\documentclass[9pt,conference]{IEEEtran}
\IEEEoverridecommandlockouts
 \usepackage{url}
\usepackage{cite}
\usepackage{amsmath,amssymb,amsfonts}
\usepackage{algorithmic}
\usepackage{graphicx}
\usepackage{textcomp}
\usepackage{booktabs} 
\usepackage{xcolor}
\def\BibTeX{{\rm B\kern-.05em{\sc i\kern-.025em b}\kern-.08em
    T\kern-.1667em\lower.7ex\hbox{E}\kern-.125emX}}
\makeatletter
\newcommand{\linebreakand}{%
  \end{@IEEEauthorhalign}
  \hfill\mbox{}\par
  \mbox{}\hfill\begin{@IEEEauthorhalign}
}
\makeatother
\begin{document}

\title{ZSVC: Zero-shot Style Voice Conversion with Disentangled Latent Diffusion Models and Adversarial Training
}

\author{
    \IEEEauthorblockN{
		\textit{Xinfa Zhu}\IEEEauthorrefmark{2}\IEEEauthorrefmark{4}, 
		\textit{Lei He}\IEEEauthorrefmark{4}, 
        \textit{Yujia Xiao}\IEEEauthorrefmark{4},
        \textit{Xi Wang}\IEEEauthorrefmark{4},
        \textit{Xu Tan}\IEEEauthorrefmark{4},
        \textit{Sheng Zhao}\IEEEauthorrefmark{4},
        \textit{Lei Xie}\IEEEauthorrefmark{2}\IEEEauthorrefmark{1}
  }
	\IEEEauthorblockA{\IEEEauthorrefmark{2}Audio, Speech and Language Processing Group (ASLP@NPU), School of Computer Science,\\ Northwestern Polytechnical University, Xian, China\\
	\IEEEauthorblockA{\IEEEauthorrefmark{4}Microsoft, Beijing, China\\
 }
}\thanks{\IEEEauthorrefmark{1}: Corresponding author. This work was done when Xinfa Zhu was an intern in Microsoft, Beijing, China.}}

\maketitle

\begin{abstract}
Style voice conversion aims to transform the speaking style of source speech into a desired style while keeping the original speaker's identity. However, previous style voice conversion approaches primarily focus on well-defined domains such as emotional aspects, limiting their practical applications. 
In this study, we present ZSVC, a novel Zero-shot Style Voice Conversion approach that utilizes a speech codec and a latent diffusion model with speech prompting mechanism to facilitate in-context learning for speaking style conversion.
To disentangle speaking style and speaker timbre, we introduce information bottleneck to filter speaking style in the source speech and employ Uncertainty Modeling Adaptive Instance Normalization (UMAdaIN) to perturb the speaker timbre in the style prompt. 
Moreover, we propose a novel adversarial training strategy to enhance in-context learning and improve style similarity.
Experiments conducted on 44,000 hours of speech data demonstrate the superior performance of ZSVC in generating speech with diverse speaking styles in zero-shot scenarios.
\end{abstract}


\section{Introduction}

Zero-shot voice conversion (VC) aims to convert speech from any source speaker to that of any target speaker without altering the linguistic content. VC operates by decomposing source speech into distinct components, including speaker timbre, linguistic content, and speaking style. Traditional zero-shot VC approaches typically merge linguistic content and the speaking style of the source speaker with the target speaker's timbre to generate converted speech~\cite{autovc,lmvc}. However, zero-shot style VC, which combines linguistic content and speaker timbre of the source speaker with the target speaker's speaking style for generating converted speech, has received limited attention.

\begin{figure}[t]
  \centering
  \includegraphics[width=\linewidth]{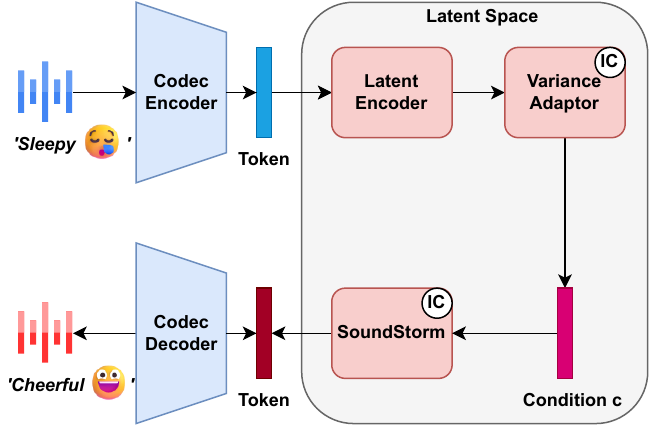}
  \caption{Overall framework of ZSVC. 'IC' means in-context learning through speech prompting mechanism}
  \label{fig:system_overview}
\end{figure}

\begin{figure*}[t]
  \centering
  \includegraphics[width=\linewidth]{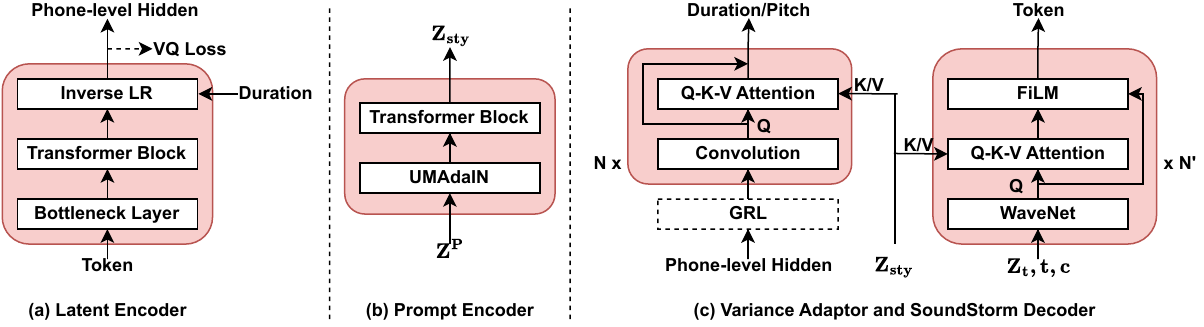}
  \caption{The detailed architecture of proposed ZSVC. The dashed line means only available in training.}
  \label{fig:system_details}
\end{figure*}

Zero-shot style VC holds significance in various scenarios, such as movie dubbing, live broadcasting, and data augmentation, offering the potential to assist individuals in expressing themselves in an appropriate speaking style. Previous style VC studies primarily concentrate on emotional aspects, which is a well-defined domain. Specifically, Variational Autoencoder (VAE)-based methods, exemplified by the hybrid VAE-GAN model~\cite{hybridvae}, have played a pivotal role in converting non-parallel emotional speech without compromising the speaker's identity and linguistic content. Advancements in emotional VC have been marked by Generative Adversarial Network (GAN)-based approaches, employing frameworks like Cycle-GAN~\cite{cyclegan}, StarGAN~\cite{stargan}, and VAEGAN~\cite{vaegan}.
Moreover, StyleVC~\cite{stylevc} proposes an any-to-any expressive voice conversion framework that decomposes speech into linguistic content, speaker identity, pitch, and style information. It leverages an auto-encoder operating on Mel-Spectrogram and minimizes mutual information to disentangle each representation.
DurFlex-EVC~\cite{DurFlexEVC} introduces a style auto-encoder and unit aligner, leveraging a diffusion model to achieve flexible and high-quality emotional voice conversion. 
Nevertheless, the human voice encompasses a broader range of expressive elements, including various style expressions beyond emotional independent nuance, such as intonation and rhythm~\cite{stylevc1,speechsplit}.


This paper focuses on zero-shot style VC, aiming to convert the speaking style of any source speech into that of a target speech while preserving linguistic content and speaker timbre. The challenges of building such a system are threefold. Firstly, it is difficult to categorize or annotate all styles of human speech, rendering style annotation unavailable. Secondly, the highly intricate entanglement between speaker timbre, linguistic content, and speaking style poses a risk of unintended style or timbre leakage~\cite{METTS}. Thirdly, many VC methods depend on global embeddings, especially sourced from speaker verification networks~\cite{autovc,xvectors,globalvc}. The representation ability of global embeddings~\cite{sefvc} constrains the zero-shot VC performance. Recent advancements in speech language models~\cite{lmvc,AudioLM,VALLE} avoid this limitation by adopting the promising in-context learning strategy, predicting the target voice based on speech prompts. However, these models are designed to clone both speaking style and speaker timbre.

To address these challenges, we propose a novel framework named Zero-shot Style Voice Conversion (ZSVC) with latent diffusion models. As illustrated in Figure~\ref{fig:system_overview}, our approach begins by utilizing a speech codec to extract speech tokens from the input speech. Speech codecs are widely employed in speech generation tasks due to their robustness and generalization~\cite{VALLE,encodec}. Subsequently, we modulate the speaking style of these speech tokens through a latent diffusion model~\cite{ns2,makeanaudio}, coupled with speech prompting mechanism to facilitate in-context learning. Finally, the modulated speech tokens are sent to the speech codec to reconstruct waveforms.
To tackle the challenge of the entanglement of various speech components, we employ an information bottleneck~\cite{BN} approach to isolate the speaking style within speech tokens; and we propose Uncertainty Modeling Adaptive Instance Normalization (UMAdaIN), a non-parametric technique, to disturb the speaker's timbre in prompted speech subtly.
Furthermore, we propose an innovative adversarial training strategy to amplify in-context learning and enhance zero-shot style similarity.
Our ZSVC model is scaled to 335 million parameters and trained on an extensive dataset of 44,000 hours of speech. The experiment results highlight ZSVC's ability to generate speech with a wide range of speaking styles in zero-shot scenarios, showcasing its potential for diverse applications.
Samples are available on our demo page\footnote{\url{https://zxf-icpc.github.io/zsvc/}}.

\section{Proposed Approach}

As depicted in Figure~\ref{fig:system_overview}, the ZSVC system includes a speech codec and a latent diffusion model. The speech codec encoder extracts speech tokens from the input speech. The speech codec decoder reconstructs the speech waveforms from speech tokens. The latent diffusion model consists of a latent encoder, a variance adaptor, and a SoundStorm~\cite {SoundStorm} decoder, jointly modulating the speaking style of the speech tokens. Notably, the speech prompting mechanism is integrated into the variance adaptor and SoundStorm decoder through a prompt encoder, enabling the control of speaking style via in-context learning.

\subsection{Disentangled latent encoder and prompt encoder} 
In general, speech tokens encapsulate almost all linguistic content, speaker timbre, and speaking style, contributing to the reconstruction of high-quality speech. Consequently, it becomes crucial to disentangle speaking style and speaker timbre in speech tokens to prevent potential leakage in style voice conversion.

As illustrated in Figure~\ref{fig:system_details}(a), the latent encoder comprises a bottleneck layer, multiple transformer blocks, and an inverse length regulator (LR). The input speech tokens traverse the speech codec codebook to acquire token embeddings. These embeddings then proceed through the bottleneck layer that forces the model to eliminate unnecessary information by projecting the embeddings into a low-dimensional space (i.e., 64-dimension). Finally, by going through the transformer blocks and the inverse LR with phone-level duration, the low-dimensional embeddings are projected onto a phone-level hidden sequence $H$.
Moreover, another information bottleneck, a vector quantization (VQ) commit loss~\cite{vqvae} is applied on $H$ to form linguistic clusters. By mapping the hidden sequence $H$ of different speaking styles to the same linguistic clusters, VQ further filters speaking style within $H$.

Additionally, inspired by the belief that instance normalization (IN) can eliminate most of the speaker timbre~\cite{IN} and drawing from the concept of Uncertainty Modeling Layer Normalization (UMLN) in StyleSinger~\cite{stylesinger}, we introduce UMAdaIN to perturb and disentangle speaker timbre in the prompt encoder, as depicted in Figure~\ref{fig:system_details}(b). Prompt speech tokens undergo the codebook to yield prompt embedding $Z^P$. We compute the mean $\mu$ and variance $\sigma$ of $Z^P$. And then we calculate the scale $\overline{\mu}$ and bias $\overline{\sigma}$ vectors through averaging $\mu$ and $\sigma$. To perturb speaker information, we employ a Gaussian distribution to model the uncertainty scope of speaker embedding. Specifically, we sample weights $\omega_1$ and $\omega_2$ from a standard Gaussian distribution. Thus the resulting speaker-agnostic hidden representation is defined as:
\begin{equation}
UMAdaIN(x) = {\omega_1 \odot{\overline{\mu} \odot {\frac{x-\mu}{\sigma}} + \omega_2 \odot \overline{\sigma}}}.
\label{eq:eq2}
\end{equation}
Finally, the speaker-agnostic hidden representation undergoes several transformer blocks to yield the style presentation $Z_{sty}$.

\subsection{SoundStorm based latent diffuion}
As depicted in Figure~\ref{fig:system_details}(c), given the phone-level hidden sequence $H$ and style presentation $Z_{sty}$, the variance adaptor predicts both duration and pitch. The hidden sequence $H$ is then expanded to the frame level based on the duration and combined with pitch embedding to form the final condition information $c$. Subsequently, the condition $c$ is forwarded to the SoundStorm decoder to predict speech tokens.

Given the condition $c$ and $Z_{sty}$, the process of predicting speech tokens $Z_{0:T}$ can be defined as follows:
\begin{align}
    p(Z_{0:T}|c,Z_{sty};\theta_s) &=\prod_{t=0}^{T}p(Z_t|Z_{<t},c, Z_{sty};\theta_s),
\end{align}
where $\theta_s$ denotes the parameters of the SoundStorm decoder and $T$ denotes the number of quantizers in the speech codec.
Specifically, we define a masking function as $m(i)$. At each time step $i$, we mask and predict a subset of $Z_t$. Therefore, we extend Equation (2) to:
\begin{equation}
\begin{aligned}  
& p(Z_{0:T}|c,Z_{sty};\theta_s) = 
\\
& \prod_{t=0}^{T}\prod_{i=0}^{n}p(m(i) \odot Z_t|(1-m(i)) \odot Z_t,Z_{<t},c, Z_{sty};\theta_s),
\end{aligned}\label{eq:eq3}
\end{equation}
where $n$ represents the total decoding steps. Notably, distinct from the original SoundStorm~\cite{SoundStorm}, we incorporate the speech prompt $Z_{sty}$ through Q-K-V attention~\cite{ns2}.

\subsection{Enhancing in-context learning with adversarial training}
We incorporate adversarial training within the variance adaptor to amplify in-context learning for optimal performance in speaking style converting. Normally, given hidden sequence $H$ and $Z_{sty}$, the variance adaptor predict duration $D$ and pitch $P$ as follows:
\begin{align}
    \mathcal{L}_{va} &= p(D,P|H,Z_{sty};\theta_{va}),
\end{align}
where $\theta_{va}$ denotes the parameters of the vaiance adaptor. During adversarial training, we eliminate $Z_{sty}$ and Q-K-V attention in the variance adaptor and introduce a Gradient Reversal Layer (GRL)~\cite{grl} on $H$, as outlined below: 
\begin{align}
   \mathcal{L}_{va'} &= p(D,P|GRL(H);\theta_{va'})
\end{align}
where $\theta_{va'}$ denotes the parameters of the modified variance adaptor with GRL. This strategy aids in removing residual speaking style in $H$ and forces the model focus $Z_{sty}$ to capture speaking style.

\subsection{Training and inference}
During training, we follow Naturalspeech 2~\cite{ns2} and use a random segment of the target speech as prompt speech, while the rest segments are used as input speech. The training objective of ZSVC is defined as: 
\begin{align}
    \mathcal{L} &= \mathcal{L}_{diff} + \mathcal{L}_{VQ} +  \mathcal{L}_{va} + \lambda_{grl}\mathcal{L}_{va'},
\end{align}
where $\mathcal{L}_{diff}$ represents the cross-entropy (CE) loss between predicted tokens and ground-truth tokens. $\mathcal{L}_{VQ}$ represents the VQ commit loss on $H$. $\mathcal{L}_{va}$ and $\mathcal{L}_{va'}$ denote the L1 loss between predicted pitch and duration and ground-truth pitch and duration, respectively. $\lambda_{grl}$ is the weight of adversarial training and is empirically set to 0.5.

In the inference phase, given the source speech and prompt speech, ZSVC synthesizes the target speech, preserving the linguistic content and speaker timbre of the source speech while incorporating the speaking style of the prompt speech.

\section{Experiment}

\subsection{Experiment setup}
We use the English subset of Multilingual LibriSpeech (MLS)~\cite{mls} as the training data containing 44K hours of transcribed speech data derived from LibriVox audiobooks. The sample rate is 16KHz for all speech data. The input text sequence is first converted into a phoneme sequence using grapheme-to-phoneme conversion~\cite{g2p} and then aligned with speech using our internal alignment tool to obtain the phoneme-level duration. Furthermore, the frame-level pitch sequence is extracted from the speech using PyWorld.

To validate the performance of ZSVC, we implement the following systems:
\begin{itemize}
    \item \textbf{LGVC}: a latent diffusion framework incorporating a global style token (GST)~\cite{gst} to control speaking style. LGVC employs the same speech codec as ZSVC to extract speech tokens. Then, the extracted speech tokens go through a transformer encoder, a variance adaptor, and a SoundStorm decoder to predict target speech tokens. The style representation extracted by GST is conditioned to the variance adaptor and SoundStorm decoder.
    \item \textbf{StyleVC}~\cite{stylevc}: an any-to-any expressive voice conversion framework that consists of a style encoder, a speaker encoder, a content encoder, and a decoder. StyleVC employs an unsupervised speech disentanglement that decomposes speech into linguistic content, speaker identity, pitch, and emotional information through mutual information minimization.
    \item \textbf{ZSVC}: the proposed zero-shot style voice conversion system with latent diffusion models.
\end{itemize}

In our implementation, the speech codec, prompt encoder, and variance adaptor follow the setting of Naturalspeech 2~\cite{ns2}. There are 6 transformer blocks with 8 attention heads, 512 embedding dimensions, a filter size of 2048, a kernel size of 9 for 1D convolution, and a dropout rate of 0.1. The bottleneck dimension is set to 64, and the number of codebooks is fixed at 8192. The diffusion model integrates 30 WaveNet layers, featuring 1D dilated convolution layers with a kernel size of 3, a filter size of 1024, and a dilation size of 2. Notably, a FiLM layer is introduced at every 3 WaveNet layers to fuse condition information processed by the Q-K-V attention in the prompting mechanism within the diffusion model. The hidden size in WaveNet is set to 512, with a dropout rate of 0.2. We set the diffusion steps to 150.
We train the models on 8 NVIDIA TESLA V100 32GB GPUs with
a batch size of 6K frames of speech tokens per GPU for 300K steps. We optimize the models with the AdamW optimizer with $5e^{-4}$ learning rate, 32k warmup steps following the inverse square root learning schedule.

\subsection{Subjective evaluation}
We conduct mean opinion score (MOS) experiments to evaluate speech naturalness, speaker similarity, and style similarity. Specifically, given 12 reserved source speech and 12 prompt speech, we generate 144 listening samples. We invite 18 volunteers to participate in the listening tests. In each test session (naturalness/speaker/style), participants are presented with randomly selected speech samples and requested to score the naturalness or similarity of each sample on a 5-point scale (`5' for excellent, `4' for good, `3' for fair, `2' for poor, and `1' for bad).

\begin{table}[htb!]
\centering
\caption{Results of subjective evaluation with 95$\%$ confidence interval.}
\label{tab:exp1}
\begin{tabular}{@{}lccc@{}}
\toprule
Model   & Naturalness ↑ & \begin{tabular}[c]{@{}c@{}}Speaker \\ Similarity ↑\end{tabular} & \begin{tabular}[c]{@{}c@{}}Style \\ Similarity ↑\end{tabular} \\ \midrule
LGVC    & \textbf{3.89 $\pm$ 0.05}               & \textbf{4.21 $\pm$ 0.06}                                                                 & 3.03 $\pm$ 0.06                                                               \\
StyleVC & 3.53 $\pm$ 0.08               & 2.89 $\pm$ 0.10                                                                 & 3.23 $\pm$ 0.09                                                               \\
ZSVC    & 3.86 $\pm$ 0.07               & 3.90 $\pm$ 0.08                                                                 & \textbf{3.96 $\pm$ 0.06}                                                               \\ \bottomrule
\end{tabular}
\end{table}


As depicted in Table~\ref{tab:exp1}, the proposed ZSVC significantly outperforms LGVC and StyleVC regarding style similarity while achieving competitive scores in naturalness and speaker similarity. LGVC excels in naturalness and speaker similarity, showcasing the advantages of the latent diffusion model. However, lacking information bottleneck and in-context learning strategy, LGVC experiences significant style leakage from the source to synthetic speech, and the target style extracted by GST is disregarded, resulting in the lowest style similarity and a failure in zero-shot style VC. Moreover, StyleVC achieves higher style similarity than LGVC but exhibits lower naturalness and speaker similarity scores. The unsupervised learning of style and speaker representations in StyleVC fails to capture desired style or speaker characteristics due to the absence of explicit constraints such as style or speaker tags. Besides, the fixed-length voice conversion in StyleVC limits its performance in terms of style similarity. ZSVC demonstrates a balanced performance, effectively addressing the challenge of speech entanglement in an unsupervised manner and successfully transferring target style through enhanced in-context learning.

\subsection{Objective evaluation}
We employ an Automatic Speech Recognition (ASR) model to transcribe the generated speech and calculate the Word Error Rate (WER) to assess the robustness of each model. The ASR model is a CTC-based Hubert, pre-trained on Librilight and fine-tuned on the 960-hour training set of LibriSpeech. We use the official code and checkpoint\footnote{\url{https://huggingface.co/facebook/hubert-large-ls960-ft}}. In addition, we employ the WavLM-TDCNN\footnote{\url{https://github.com/microsoft/UniSpeech/tree/main/downstreams/speaker\_verification}}
speaker embedding model to assess speaker similarity (SSIM)
between generated samples and source speech. Moreover, for style similarity, we employ two pitch-related metrics Root Mean Squared Error (RMSE) and Pearson correlation (Corr), and take the emotional speech dataset (ESD)~\cite{esd} as the test set. These two metrics are widely applied to evaluate the prosody similarity of VC. Since the sequences are not aligned, we perform Dynamic Time Warping to align the sequences before comparison. Besides, we measure the RMSE/Corr between source speech and generated speech as S-RMSE/S-Corr, while the RMSE/Corr between target speech and generated speech is denoted as T-RMSE/T-Corr.
\begin{table}[htb]
\centering
\caption{Results of objective evaluation}
\setlength\tabcolsep{3pt}
\label{tab:exp2}
\begin{tabular}{@{}lccccccc@{}}
\toprule
Model   & WER ↓ & SSIM ↑ & S-RMSE ↑ & S-Corr ↓ & T-RMSE ↓ & T-Corr ↑ \\ \midrule
LGVC    & \textbf{1.98}         & \textbf{0.934}    & 3.34       & 0.95       & 43.61      & 0.56       \\
StyleVC & 5.52         & 0.624    & 10.62      & 0.87       & 36.93      & 0.65       \\
ZSVC    & 2.34         & 0.856    & \textbf{36.47}      & \textbf{0.59}       & \textbf{12.11}      & \textbf{0.88}       \\ \bottomrule
\end{tabular}
\end{table}

Objective results are presented in Table~\ref{tab:exp2}. LGVC achieves the lowest WER and highest SSIM, indicating faithful reconstruction of the source speech in synthetic speech. This observation aligns with the prosody similarity results, where LGVC exhibits the closest prosody similarity (lowest S-RMSE and highest P-Corr) to source speech but deviates significantly from prompt speech (highest P-RMSE and lowest P-Corr). StyleVC exhibits a higher WER and lower SSIM, highlighting challenges in traditional autoencoder frameworks for zero-shot style VC. However, our proposed approach, ZSVC, incorporating speech codec and latent diffusion models, achieves comparable WER and SSIM, showcasing high robustness, intelligibility, and speaker similarity. Furthermore, ZSVC demonstrates the highest prosody similarity with prompt speech and a notable divergence from source speech, underscoring the efficacy of speech disentanglement and style transfer through enhanced in-context learning.

\subsection{Visual Analysis}
\begin{figure}[htb]
\centering
\begin{minipage}[b]{0.49\linewidth}
  \centering
  \includegraphics[width=\textwidth]{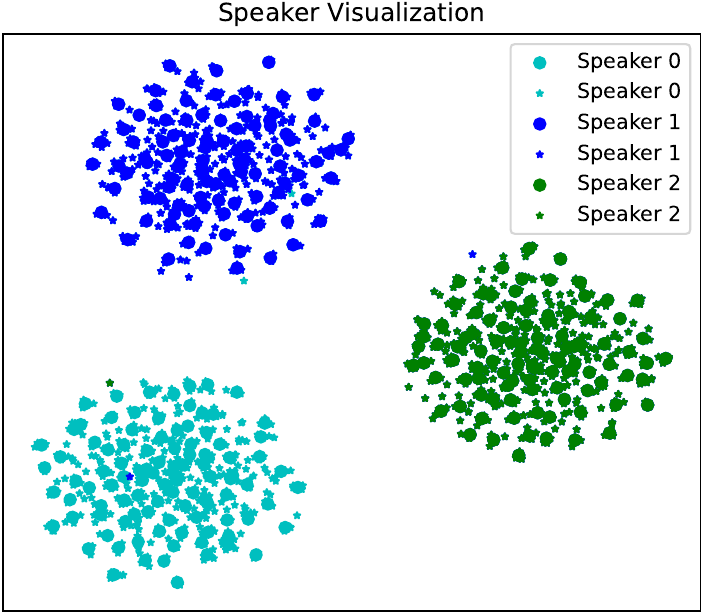}
\end{minipage}
\hfill
\begin{minipage}[b]{0.49\linewidth}
  \centering
  \includegraphics[width=\textwidth]{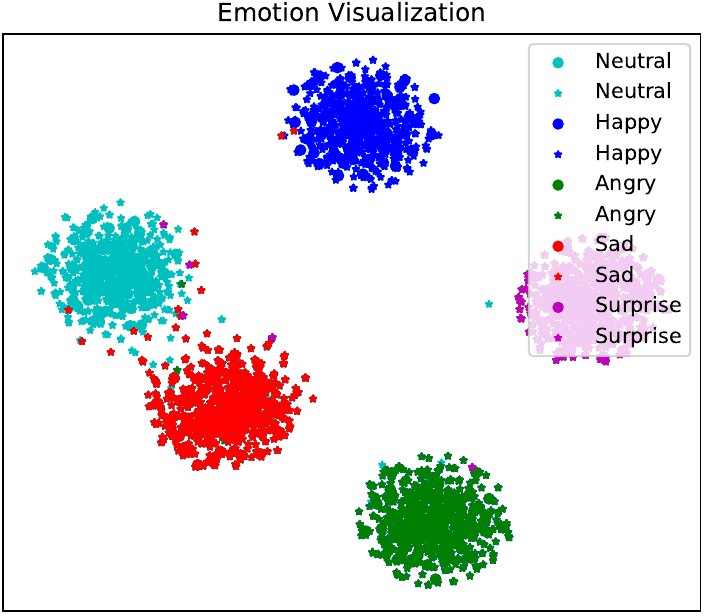}
\end{minipage}

\caption{T-SNE visualization of speaker representations (left) and emotion representations (right). The circle represents the original speech, and the star represents the converted speech.}
\label{fig2}
\end{figure}

We further analyze the speaker similarity and prosody variance of converted speech through T-SNE~\cite{tsne}. Specifically, we randomly select 100 audio samples from five speakers as source speech and 5 audio samples from other speakers as prompt speech to conduct zero-shot style voice conversion via ZSVC.
We use WavLM-TDCNN and emotion2vec~\cite{emo2vec} to extract speaker and emotion embeddings, respectively. After that, we visualize them before and after style conversion within one subfigure. As depicted in Figure~\ref{fig2}, which presents clear clusters. This observation validates that ZSVC maintains the speaker timbre of source speech while converting speaking style.

\subsection{Ablation study}
We conduct an ablation study to validate the key components of ZSVC by removing certain modules. The corresponding results are presented in Table~\ref{tab:exp3}.  The removal of the information bottleneck (-BN) significantly impacts the naturalness and style similarity of synthetic speech. The absence of information bottlenecks results in the leakage of abundant undesired information, such as the style of the source speech. This leakage burdens the adversarial training process in disentangling speaking style and makes the model unstable.
Furthermore, removing UMAdaIN (-UMAdaIN) leads to the leakage of the speaker timbre of the prompt speech into synthetic speech, resulting in a notable decline in speaker similarity.
Additionally, when adversarial training is omitted (Adversarial Training), naturalness and style similarity decrease. This suggests that the in-context learning enhanced by adversarial training not only improves style transfer but also contributes to achieving natural speech.
\begin{table}[htb!]
\centering
\caption{Results of ablation study with 95$\%$ confidence interval.}
\label{tab:exp3}
\setlength\tabcolsep{5pt}
\begin{tabular}{@{}lccc@{}}
\toprule
Model   & Naturalness ↑ & \begin{tabular}[c]{@{}c@{}}Speaker \\ Similarity ↑\end{tabular} & \begin{tabular}[c]{@{}c@{}}Style \\ Similarity ↑\end{tabular} \\ \midrule
ZSVC    & \textbf{3.86 $\pm$ 0.07}               & \textbf{3.90 $\pm$ 0.08}                                                                 & \textbf{3.96 $\pm$ 0.06}                                                               \\
- BN & 3.58 $\pm$ 0.10               & 3.82 $\pm$ 0.08                                                                 & 3.40 $\pm$ 0.10                                                               \\
- UMAdaIN & 3.80 $\pm$ 0.07               & 3.61 $\pm$ 0.06                                                                 & 3.89 $\pm$ 0.06                                                               \\
\begin{tabular}[c]{@{}c@{}}- Adversarial \\ Training\end{tabular}    & 3.77 $\pm$ 0.05               & 3.90 $\pm$ 0.06                                                                 & 3.77 $\pm$ 0.05                                                               \\ \bottomrule
\end{tabular}
\end{table}
\section{Conclusion}

This paper introduces ZSVC, a novel approach for zero-shot style voice conversion. ZSVC harnesses the power of speech codec and latent diffusion models, utilizing a speech codec to extract speech tokens and employing the latent diffusion model with speech prompt mechanism to modulate the speaking style of these tokens. To address the challenge of speech entanglement, we introduce an information bottleneck to filter speaking style in source speech and utilize Uncertainty Modeling Adaptive Instance Normalization (UMAdaIN) to perturb the speaker timbre in prompt speech. Furthermore, we propose a novel adversarial training strategy to enhance in-context learning and improve style similarity.
We conducted extensive experiments on 44,000 hours of multilingual libriSpeech. The experimental results underscore ZSVC’s superiority in generating speech with diverse speaking styles in zero-shot scenarios.

\clearpage
\bibliographystyle{IEEEbib}
\bibliography{refs}

\end{document}